\documentclass[times]{weauth}
\usepackage{moreverb}
\usepackage{subfigure}
\usepackage{graphicx}
\usepackage{dcolumn}
\usepackage{amssymb,amsmath}
\usepackage{epstopdf}
\usepackage{placeins}
\usepackage{citesort}
\usepackage{color}

\def\volumeyear{2014}

\begin{document}

\runningheads{Stevens, Gayme, Meneveau}{Effects of turbine spacing on the power output of extended wind-farms}
\articletype{\noindent "This is the peer reviewed version of the following article: \textbf{Stevens, R. J. A. M., Gayme, D. F., and Meneveau, C. (2016) Effects of turbine spacing on the power output of extended wind-farms. Wind Energ., 19: 359-370}, which has been published in final form at \textbf{http://dx.doi.org/10.1002/we.1835}. This article may be used for non-commercial purposes in accordance with Wiley Terms and Conditions for Self-Archiving." \nobreak}
\title{Effects of turbine spacing on the power output of extended wind-farms}
\author{Richard J.A.M. Stevens$^{1,2}$, Dennice F. Gayme$^{1}$, and Charles Meneveau$^{1}$}
\address{$^1$ Department of Mechanical Engineering, Johns Hopkins University, Baltimore, Maryland 21218, USA\\
$^2$ Department of Physics, Mesa+ Institute, and J.\ M.\ Burgers Centre for Fluid Dynamics, University of Twente, 7500 AE Enschede, The Netherlands}
\corraddr{r.j.a.m.stevens@jhu.edu}

\begin{abstract}
We present results from large eddy simulations (LES) of extended wind-farms for several turbine configurations with a range of different spanwise  and streamwise  spacing combinations. The results show that for wind-farms arranged in a staggered configuration with spanwise spacings in the range $\approx[3.5,8]$D, where $D$ is the turbine diameter, the power output in the fully developed regime depends primarily on the geometric mean of the spanwise and streamwise turbine spacings. In contrast, for the aligned configuration the power output in the fully developed regime strongly depends on the streamwise turbine spacing and shows weak dependence on the spanwise spacing. Of interest to the rate of wake recovery, we find that the power output is well correlated with the vertical kinetic energy flux, which is a measure of how much kinetic energy is transferred into the wind-turbine region by the mean flow. A comparison between the aligned and staggered configurations reveals that the vertical kinetic energy flux is more localized along turbine rows for aligned wind-farms than for staggered ones. This additional mixing leads to a relatively fast wake recovery for aligned wind-farms.\\
\end{abstract}


\maketitle

\section{Introduction}
It is well known that wakes created by upstream wind-turbines can significantly reduce the power production of downstream turbines and that this phenomenon can significantly affect wind-farm performance. Improving our fundamental understanding of wake effects and overall power extraction efficiency is critical for the optimization of increasingly larger wind-farms. Considering standard wind-farms with turbines arranged on a lattice, the spanwise and streamwise turbine spacings are design parameters and it is important to know how the power output depends on these spacings. Another important parameter is the geometric mean turbine spacing, which is defined as $s=\sqrt{s_xs_y}$, where $s_x$ and $s_y$ indicate the streamwise and spanwise dimensionless spacings (in units of rotor diameter) respectively. Intuitively,  it is to be expected that the spanwise and streamwise spacing have a different influence on the turbine power output. In smaller wind-farms, or in the entrance region of large wind-farms, the power output of the turbines for a fixed geometric mean turbine spacing indeed strongly depends on the combination of the spanwise and streamwise turbine spacing that is chosen. Such trends are borne out, e.g., in studies using engineering wake models \cite{new77,lis79,mil80,fra92,fra06}. Studies that have addressed the effect of the streamwise and spanwise spacing in the fully developed regime, i.e., for wind-farms that are sufficiently extended so that the turbine power output becomes independent of the downstream position, are more limited. It is, for example, known that engineering wake models have difficulty in predicting wake effects in the fully developed regime, which is likely due to their inability to capture the time-dependent interaction of multiple turbine wakes correctly \cite{bar09b,son14}. A recent illustration of the limitations of wake models in predicting power output in the deep-array limit of staggered wind-farms is provided in Ref. \cite{ste14g}. These engineering models use a simple, linear equation to rapidly calculate wakes for the thousands of instances required to find the most efficient turbine configuration \cite{jen84,kat86,nyg14}. More advanced models exist, a subset of these are based on a parametrization of the internal boundary layer growth coupled with some eddy viscosity model (e.g.\ the Deep-Array Wake Model \cite{bro12} and the Large Array wind-farm model \cite{has09}). Other models such as FUGA \cite{ott11}, Windmodeller \cite{bea12}, Ellipsys \cite{iva08},  Reynolds-Averaged Navier Stokes (RANS) methods (Ainslie model \cite{ain88}), UPMPARK, WakeFarm \cite{pij06,sch07b} and Farmflow \cite{eec10,ozd13} estimate the wake using a linearized computational fluid dynamics (CFD) model. A description and review of a number of these methods is provided in Ref.\ \cite{san11}.

To achieve more reliable results with less reliance on modeling assumptions, high-fidelity computer simulations of wind-farms have gained attention as a tool for studying flow in wind-turbine arrays, and, in particular, wake effects. The tool that requires the least modeling assumptions is direct numerical simulation (DNS). However, DNS is not a computationally tractable method for wind-farm simulations because of the very large scale disparities associated with high Reynolds number flows in the atmospheric boundary layer (ABL) and the near-blade boundary layers. Large eddy simulations (LES), which have intermediate complexity between DNS and the engineering models, provide an appealing alternative means of studying wake interactions in wind-farms. LES are similar to a DNS in that the largest scales of the flow are fully space and time resolved, however they are more computationally tractable because the effect of the smallest scales are represented by a subgrid-scale model. Recently it has been shown that LES can accurately capture wake interactions and kinetic energy transport in large-wind-farms \cite{cal10,cal11,wu11,lee12,chu12b,yan12,wu13,wu13b,arc13b,por13,yan14,yan14b,ste14,ste14b}. More background on LES for wind-farm studies can be found in Mehta et al.\ \cite{meh14}. In the study of Yang et al.\ \cite{yan12} LES are used to study the effect of the streamwise and spanwise spacing on the power output in the fully developed regime for aligned wind-farms. They showed that, for a given geometric mean turbine spacing, a higher power output is obtained when the streamwise spacing is larger. Their results imply that the streamwise spacing is more important than the spanwise spacing in the design of aligned wind-farms. This result from Yang et al.\ \cite{yan12} is in contrast to assumptions made in some theoretically derived models that predict the effective roughness height of large wind-farms \cite{fra92,fra06,cal10,men12,ste14c}. In these models, the horizontally averaged mean velocity at hub-height and the corresponding predicted power output of the turbines depends on this effective roughness height because it is a measure of the mean velocity that is obtained at hub-height. Due to the use of a horizontally averaged mean velocity the effect of the turbine positioning is not explicitly included in these models and the predicted effective roughness height only depends on the geometric mean turbine spacing. It has been shown by Stevens et al.\ \cite{ste14g} that the top-down model better represents the flow properties in a staggered wind-farm than in an aligned wind-farm.

\begin{figure}
\centering
\includegraphics[width=0.50\textwidth]{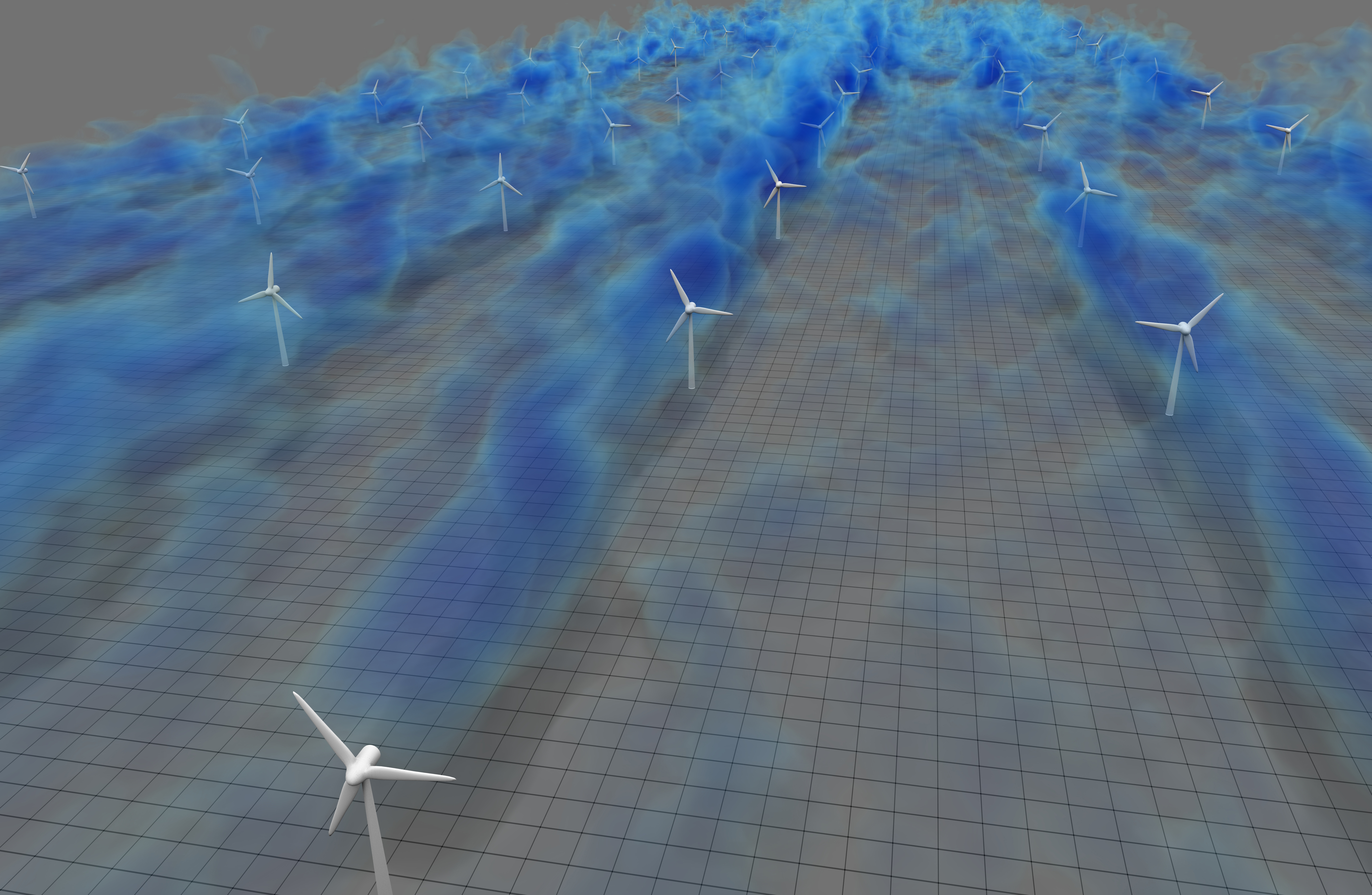}
\caption{A three-dimensional visualization of the flow in the simulated wind-farm. The figure shows a density plot of the low streamwise velocity (blue regions) found mainly behind the turbines. Turbines are represented in LES using the actuator disk model and are shown with individual blades in the figure solely for illustration purposes. The spatial resolution of the pseudo-spectral method in horizontal planes is indicated by the grid on the bottom surface. (Visualization courtesy of David Bock (NCSA, National Center for Supercomputing Applications and XSEDE, Extreme Science and Engineering Discovery Environment) as part of the Extended Collaborative Support Services of XSEDE).}
\label{figure1}
\end{figure}

It is known that a larger streamwise turbine spacing and using a staggered turbine configuration instead of an aligned one is beneficial for the power output in the fully developed regime of the wind-farm. However, the interplay between these important wind-farm design parameters is insufficiently understood. The first step in developing an improved understanding is to acquire reliable data describing the wind-farm performance as function of the streamwise and spanwise spacing, and relative turbine placement. In order to obtain such data we use high-fidelity LES to study wind-farm performance for different combinations of spanwise and streamwise turbine spacing and several relative turbine configurations. In section \ref{section_LES} we will introduce the LES framework used in this study. Subsequently, we will discuss the main results of the simulations in section \ref{section_Results}. Here, we will first present the results for the power output, focusing on the fully developed regime of the wind-farms. In agreement with the result from Yang et al.\ \cite{yan12} we find that for an aligned wind-farm the streamwise spacing between the turbines is the most important parameter for determining power output. However, for a staggered configuration we find that it is the geometric mean turbine spacing that has the dominant impact on power output. The flow properties in a staggered configuration are more consistent with the assumptions of effective roughness height models \cite{fra92,fra06,cal10,men12,ste14c}, such as horizontal homogeneity of the flow.  This new data thus reveals important trends that should be captured in wind-farm design tools. To further improve our understanding of the observed phenomena we analyze the relation between the vertical kinetic energy flux and the power output density in the fully developed regime for the different simulations. At the end of the paper we summarize the main conclusions and give a short outlook to future work. 

\section{Large eddy simulations framework} \label{section_LES}
We model wind-farms consisting of a regular array of wind-turbines, each having a diameter of $D=100$ m and a hub-height of $z_H=100$ m. The computational domain employed in most cases is $25.14 \times 3.14 \times 2$ km ($L_x\times L_y \times L_z$) in the streamwise, spanwise and vertical directions, respectively. In addition, two cases are calculated on a $37.71\times3.14\times2$ km domain to verify that the fully developed regime, which is the regime in which the power output as function of the downstream position becomes constant, is reached for all cases. We consider wind-farms with more than ten rows in the streamwise direction in order to study the fully developed state.  This number of turbine rows assures that the power output of the later rows is approximately constant in both the staggered and aligned wind-farm configurations. The mean inflow is in the streamwise ($x$) direction for all cases. The distances between wind-turbines are $s_xD$ and $s_yD$ in the streamwise and spanwise directions, respectively. We describe the overall dimensionless turbine spacing by the geometric mean $s=\sqrt{s_x s_y}$. We vary the streamwise $s_x$ and spanwise $s_y$ spacings in the range $\sim[3.49,7.85]$ and consider different combinations of streamwise and spanwise spacings with the same geometric mean turbine spacing. These specific values arise from the use of particular domain sizes ($4 \pi  \times \pi$ km and $6\pi \times \pi$ km), in the streamwise and spanwise direction respectively, where wind-turbines are placed to allow easier comparison with earlier results. We also use $s_x/s_y$ ratios consistent with our previous work to simulate wind-farms using periodic boundary conditions \cite{cal10,ste13,ste14e}. In addition, we adjust the wind-farm layout by adjusting the angle $\psi=\arctan({s_{dy} /s_x})$ with respect to the incoming flow direction, where $s_{dy}$ indicates the spanwise offset from one turbine row to the next \cite{ste14b}. Thus $\psi=0$ degrees corresponds to an aligned wind-farm. The alignment angle corresponding to the staggered arrangement is given in table \ref{table1} as $\psi_{\footnotesize{\mbox{max}}}$. In this paper we focus on the aligned and staggered configurations and show results for intermediate $\psi$ only for a few selected cases, see table \ref{table1}. The inflow in our simulations is obtained using a concurrent-precursor method \cite{ste14}. The turbines are represented by an area averaged actuator disk model \cite{jim07,cal10,yan13,ste14} using a constant thrust coefficient $C_T$, which is representative of turbines operating in regime II. Further details about the simulations can be found in Ref.\ \cite{ste14}. In order to verify that the grid resolution is not influencing our main findings we perform simulations using different grid resolutions. In the appendix we show some comparisons of different resolution simulations to show that the power output is reasonably well converged for the numerical resolutions employed in this study. The results in the remainder of this paper correspond to the $*3$ cases, where $*$ indicates the simulation cases A-H in table \ref{table1}.

Our LES code has  been validated against similar LES codes as is described by Calaf et al.\ \cite{cal10} and Yang et al.\ \cite{yan14}, which show that good agreement among these codes is obtained. In addition, Yang et al.\ \cite{yan14} and Wu and Port\'e-Agel \cite{wu13} showed that the wake profiles obtained from such a simulation approach agree well with wind tunnel experiments after a distance of $3D$ into the wake. Also, it was recently found \cite{ste13,meh14} that LES show closer agreement to the deep wake effects in large wind-farms such as Horns Rev than some engineering wake models. LES can therefore be very valuable for the further improvement of wind-farm design tools.

Figure \ref{figure1} shows a visualization for case $A3$ in table \ref{table1}. The figure shows qualitatively that the simulations capture the main wake properties and the interactions between the different wakes further downstream in the wind-farm. However, due to the large scale of the wind-farm (the large size of wind farms considered in this study was the main motivation for using an actuator disk model), the grid spacing is limited such that effects like tip vortices cannot be captured. 

\begin{table}
\caption{Table of the considered LES cases. It lists the case name, the numerical resolution in the streamwise, spanwise, and vertical direction ($N_x\times N_y \times N_z$), the streamwise ($s_x$) and spanwise ($s_y$) spacing between the turbines (in units of rotor diameter), the geometric mean turbine spacing $s=\sqrt{s_xs_y}$, the alignment angles considered ($\psi$ (degrees)) and the maximum alignment angle $\psi_{\footnotesize{\mbox{max}}}$, i.e. the angle corresponding to the staggered case. Except for case $H3$, which is calculated on a $37.71\times3.14\times2$ km domain the considered domain size is $25.14\times3.14\times2$ km.}
\label{table1}
\begin{center}
\begin{tabular}{|c|c|c|c|c|c|c|c|}
\hline
Case	 & $N_x\times N_y \times N_z$ & $s_x \times s_y$	&$s=\sqrt{s_x s_y}$		& $\psi_{\footnotesize{\mbox{max}}}$ 		& $\psi$ (degrees) 		 \\ \hline	
A3	&$1024\times128\times256$ & $7.85 \times 5.24$	& 6.41		& 	18.43								& $0.00; 3.81; 7.59; 9.46; 11.31; 14.93; 18.43$ 		 \\ \hline
B3	&$1024\times128\times256$ & $7.85 \times 3.49$	& 5.24		&	12.53								& $0.00; 3.38; 6.77; 9.46; 11.31; 12.53$ 			 \\ \hline
C3	&$1024\times128\times256$ & $5.24 \times 5.24$	& 5.24		& 	26.57								& $0.00; 2.86; 5.71; 8.53; 14.04; 19.29; 26.57$		 \\ \hline
D3	&$1024\times128\times256$ & $5.24 \times 3.49$	& 4.28		&	18.45								& $0.00; 18.45$							 \\ \hline
E3	&$1024\times128\times256$ & $5.24 \times 7.85$	& 6.41		&	36.89								& $0.00; 4.29; 8.54; 12.69; 20.57; 27.72; 36.89$ 	 \\ \hline
F3	&$1024\times128\times256$ & $3.49 \times 7.85$	& 5.24		& 	48.36								& $0.00; 48.36$							 \\ \hline
G3	&$1024\times128\times256$ & $3.49 \times 5.24$	& 4.28		& 	36.89								& $0.00; 36.89$							 \\ \hline
H3	&$1536\times128\times256$ & $7.85 \times 7.85$	& 7.85		&      26.57								& $0.00; 26.57$							 \\ \hline
A4	&$1536\times192\times384$ & $7.85 \times 5.24$	& 6.41		& 	18.43								& $0.00$									 \\ \hline
B4	&$1536\times192\times384$ & $7.85 \times 3.49$	& 5.24		&	12.53								& $0.00; 12.53$ 							 \\ \hline
C4	&$1536\times192\times384$ & $5.24 \times 5.24$	& 5.24		& 	26.57								& $0.00; 26.57$							 \\ \hline
 \end{tabular}
\end{center}
\end{table}

\section{Results} \label{section_Results}
In this section we compare the effects of the streamwise spacing, the spanwise spacing, and the geometric mean turbine spacing on the overall power output of a wind-farm. We begin in section \ref{section_power1} by discussing aligned and staggered turbine arrangements. In section \ref{section_power2} we consider intermediate turbine arrangements for a subset of the cases. We then discuss the resulting vertical kinetic energy flux and wake recovery for the aligned and staggered configurations in section \ref{section_flux}.

\subsection{Power output aligned and staggered} \label{section_power1}

\begin{figure}
\centering
\includegraphics[width=0.995\textwidth]{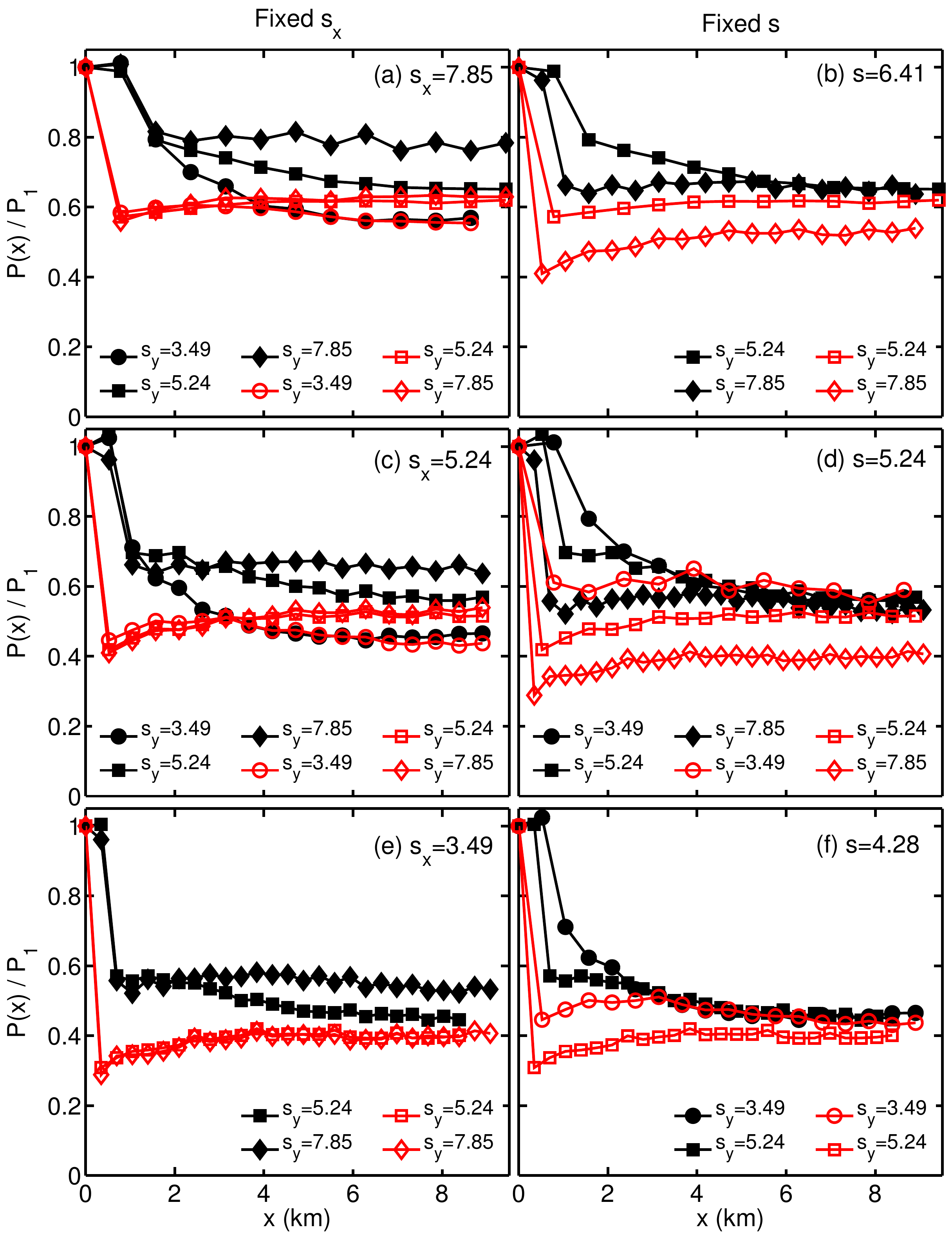}
\caption{The normalized power $P(x) / P_1$ as function of the downstream position $x$. The left panels indicate the results for a fixed streamwise spacing $s_x$ of (a) $s_x=7.85$, (c) $s_x=5.24$ and (e) $s_x=3.49$. The right panels indicate the results for a fixed geometric mean turbine spacing $s=\sqrt{s_xs_y}$ of (b) $s=6.41$, (d) $s=5.24$, and (f) $s=4.28$. Open red and solid black symbols indicate the results for the aligned and staggered configurations, respectively. }
\label{figure2}
\end{figure}

Figure \ref{figure2} shows the normalized turbine power output $P(x) / P_1$, averaged per row (streamwise location), for the different cases presented in table \ref{table1} as function of the downstream position. The left panels of figure \ref{figure2} show the results for a constant streamwise spacing and the right panels for a constant geometric mean turbine spacing $s$. The open red and solid black symbols indicate the results for the aligned and staggered cases respectively. This figure shows some interesting trends. Panels (a), (c) and (e) show that the relative turbine power output as function of the downstream position develops almost identically for the aligned configurations with spanwise spacings of $7.85$D and $5.24$D. Thus, for an aligned wind-farm we may conclude that increasing the spanwise spacing beyond $\sim 5$D has no beneficial effect on the power output. A comparison of the results for the spanwise spacings of $3.49$D and $5.24$D shown in panels (a) and (c) indicates that differences between these cases arise at a downstream distance of about $3$ km. At that point the turbine power output in the simulation with a spanwise spacing of $3.49$D becomes smaller than the power output obtained with $s_y=5.24$. The left panels of figure \ref{figure2} also show that for the staggered configuration the power output in the fully developed regime also depends significantly on the spanwise spacing. Moreover, the right panels of figure \ref{figure2} reveal that in the fully developed regime the power output for staggered wind-farms mainly depends on the geometric mean turbine spacing. In addition, the results in panels (b), (d) and (f) demonstrate that for a fixed geometric mean turbine spacing the power output in the entrance region depends strongly on both the streamwise and spanwise turbine spacings. Specifically, these results demonstrate that the turbine power output reduces more gradually when the streamwise spacing is larger. 

\begin{figure}
\centering
\subfigure{\includegraphics[width=0.49\textwidth]{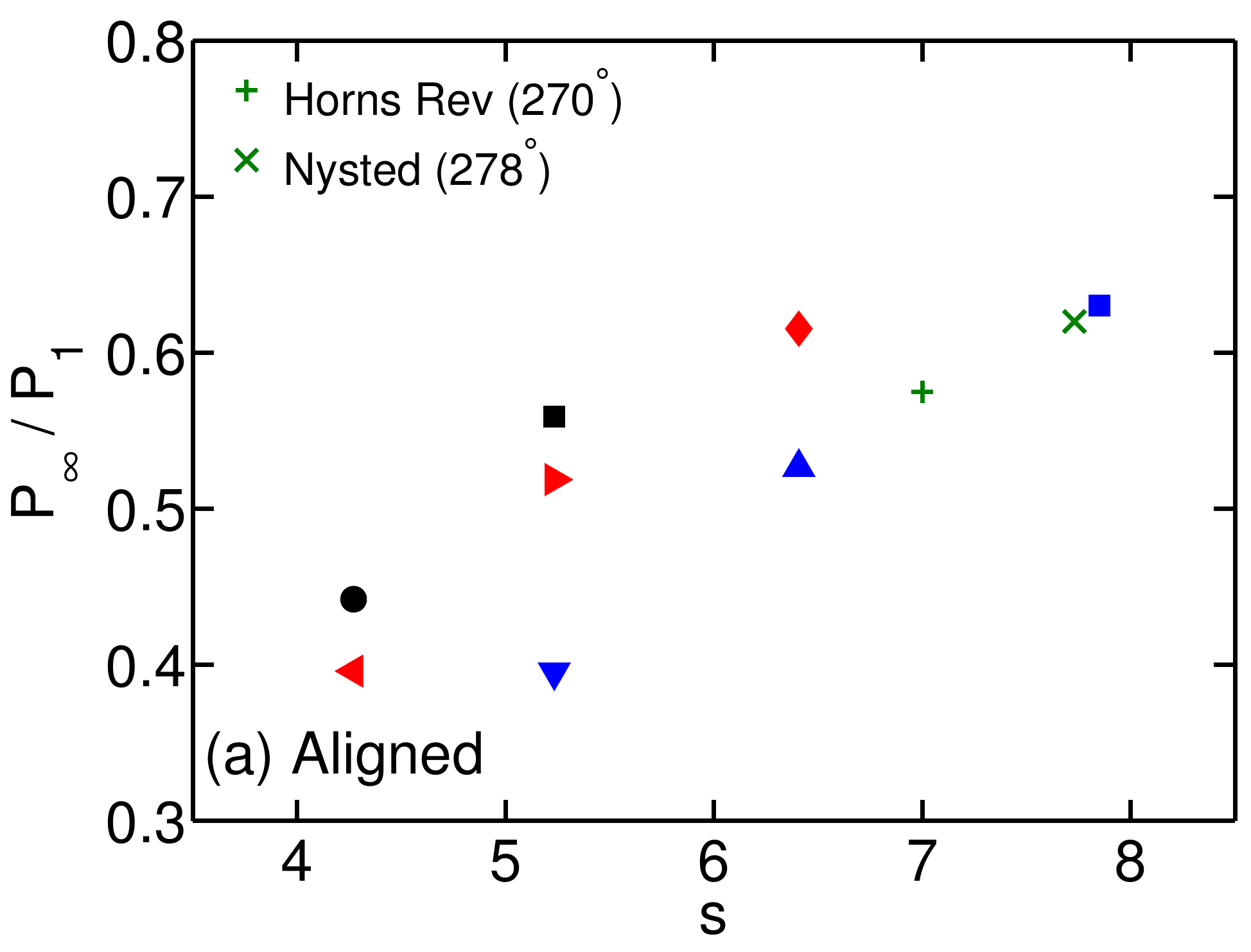}}
\subfigure{\includegraphics[width=0.49\textwidth]{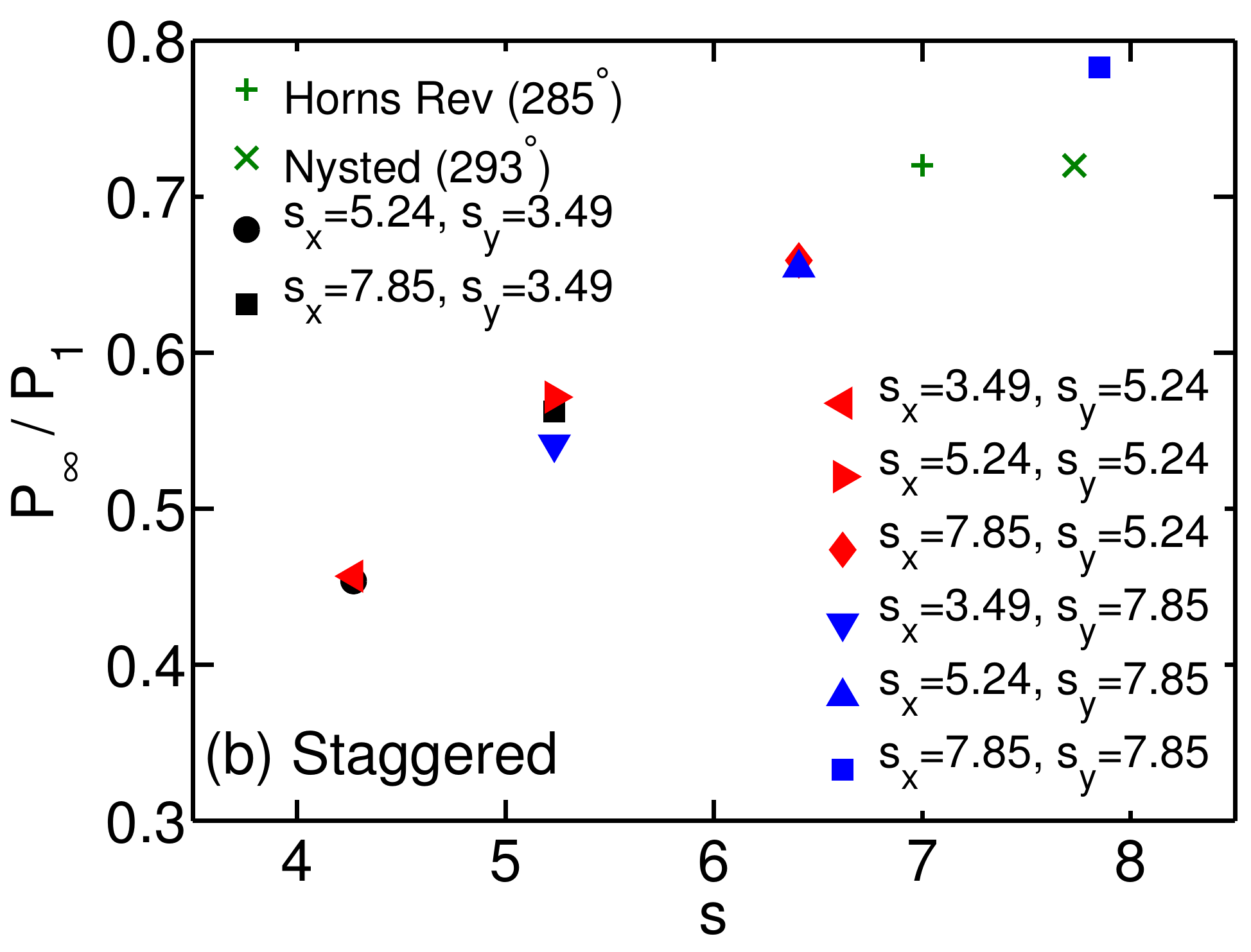}}
 \caption{Normalized power output in the fully developed regime of the wind-farms $P_{\infty}/P_1$ as function of the geometric mean turbine spacing $s=\sqrt{s_xs_y}$ for the (a) aligned and (b) the staggered configuration. Field measurement data from Horns Rev and Nysted \cite{bar11,bar09c} are included in both panels for comparison. The symbols have the same meaning in both panels.}
\label{figure3}
\end{figure}

Figure \ref{figure2} shows that the power output for the different cases remains approximately constant at locations greater than $6$ km downstream, and for several configurations this constant power output as function of the downstream position is observed closer to the wind-farm entrance. To isolate the power output in the fully developed regime for the staggered and the aligned configurations we have averaged the relative power output in the region occurring from $6$ to $8$ km downstream of the wind-farm entrance. We selected $6$ km as the start of the region based on the observation from figure \ref{figure2} that the turbine power output as function of the downstream position is nearly constant after this point and the $8$ km as the end of the region in order to ensure that this region ends within the wind-farm for all of the cases considered. Here we take the average to get a consistent comparison with the vertical flux measurements that will be presented in section \ref{section_flux} and to reduce the scatter in the data. Figure \ref{figure3} shows the corresponding relative power output, which we denote $P_\infty/P_1$, as function of the geometric mean turbine spacing for the different cases. This figure confirms that for the staggered configuration the power output in the fully developed regime mainly depends on the geometric mean turbine spacing, while for the aligned configuration it mainly depends on the streamwise distance between the turbines. We note that there are likely to be second-order effects of smaller magnitudes, but these will not be discussed here. 

In figure \ref{figure3} we compare the simulation results to field measurement data for the relative power output in the fully developed regime from Horns Rev and Nysted \cite{bar09c,bar11}. In Horns Rev the streamwise and spanwise distance between the turbines is $7D$, while Nysted has a streamwise spacing of $10.3D$ and a spanwise spacing of $5.8D$. There is data for both of these wind-farms corresponding to an aligned configuration~\cite{ste13} but neither of these wind-farms has field data that directly corresponds to a staggered configuration. So, for the comparisons with the LES of staggered wind-farms we select wind-directions of $285^{\circ}$ for Horns Rev and $293^{\circ}$ for Nysted as these conditions most closely approximate a staggered arrangement. Figure \ref{figure3}b shows that the LES and these field measurements are also in close agreement.

\begin{figure}
\centering
\subfigure{\includegraphics[width=0.70\textwidth]{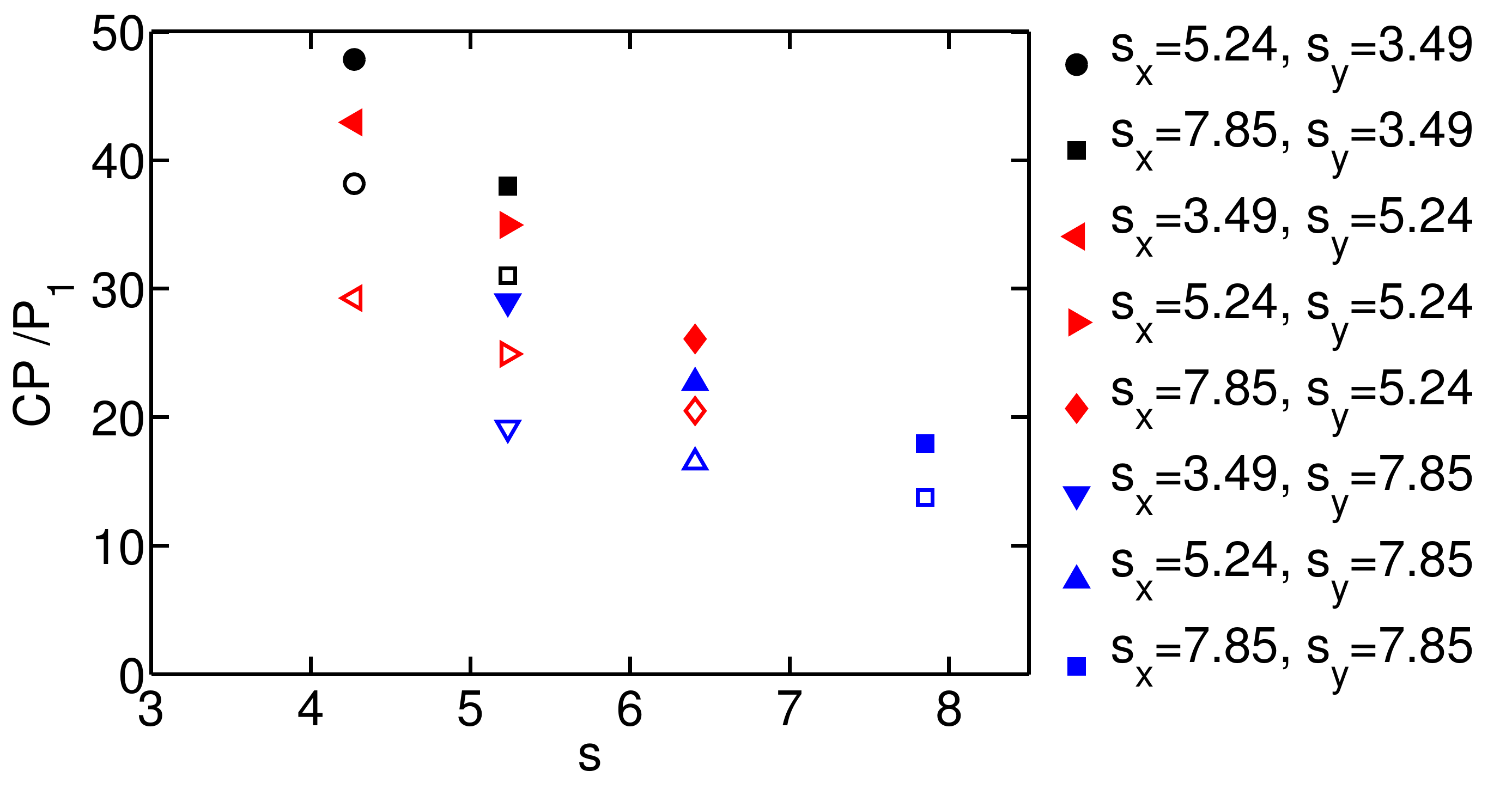}}
 \caption{Normalized cumulative power $CP / P_1$ for the the entrance region of the wind-farms, i.e.\ the $12.57$(km)$^2$ region corresponding to a streamwise distance of $4$km starting at the entrance of the wind-farm, as function of the geometric mean turbine spacings $s=\sqrt{s_xs_y}$ for the aligned (open symbols) and staggered (solid symbols) configurations. }
\label{figure4}
\end{figure}

We now shift our attention from the fully developed regime to the wind-farm entrance region. In order to investigate the entrance effects we have computed the cumulative power output of all wind-farms up to a distance of $4$km downstream (which we consider the ``entrance region''). The cumulative power for this $12.57$(km)$^2$ region is obtained by interpolating the cumulative power output corresponding to a $4$km long wind-farm. Figure \ref{figure4} shows the cumulative power output of the different wind-farms normalized by the power output of one free standing turbine at the first row, which we denote $CP/P_1$. The magnitude $CP$ depends on the chosen reference area of $12.57$(km)$^2$, but the relative values $CP/P_1$ do not depend on this choice.

The results in figure \ref{figure4} show that the power output of the wind-farm is highest when a staggered configuration with a small inter-turbine spacing, $s$, is used. In the remainder of the paper we will use $s$ and the geometric mean turbine spacing $s$, interchangeably. Wind-farms with a small inter-turbine spacing have the highest output because the loss in turbine performance due to the smaller inter-turbine spacing is more than compensated for by the larger number of turbines per unit area. However, one has to keep in mind that a wind-farm designer typically minimizes the cost of energy rather than maximizing the energy production per land area. Thus, the spacing used in actual wind-farms is closer to $7$D for both the spanwise and streamwise directions \cite{mey12,ste14c}. An interesting comparison is the aligned case with $s_x = 7.85$ and $s_y = 3.49$ compared to the staggered wind-farm case with $s_x = 3.49$ and $s_y = 7.85$. Figure \ref{figure4} shows that this aligned configuration gives a higher power output than this staggered configuration even though the turbine density and the distance between turbines that are directly upstream from one another is the same for both cases, i.e. for the staggered configuration the distance between turbines that are directly upstream from one another is based on two rows, so this distance is $2\times3.49=7.85D$. The difference between both cases is the distance measured with respect to turbines in ``neighboring columns''. For the aligned case the distance between a reference turbine and the turbines in the neighboring columns is 3.49D in the spanwise direction and 7.85D, 15.7D, 23.55D, etc. in the streamwise direction. For the staggered configuration the next neighboring column is also 3.49D away ($s_y=7.85/2=3.49$ because of the staggered arrangement) but the streamwise distance between neighboring columns is smaller than in the aligned case. For the staggered case, the sideways upstream turbines are at streamwise distances of 3.49D, 11,34D, 19.19D, etc. Therefore, partial wake interactions are slightly more important for this particular staggered case as compared to this aligned case, leading to slightly higher power output for the aligned arrangement.

\subsection{Power output for intermediate wind-turbine alignments} \label{section_power2}

\begin{figure}
\centering
\subfigure{\includegraphics[width=0.99\textwidth]{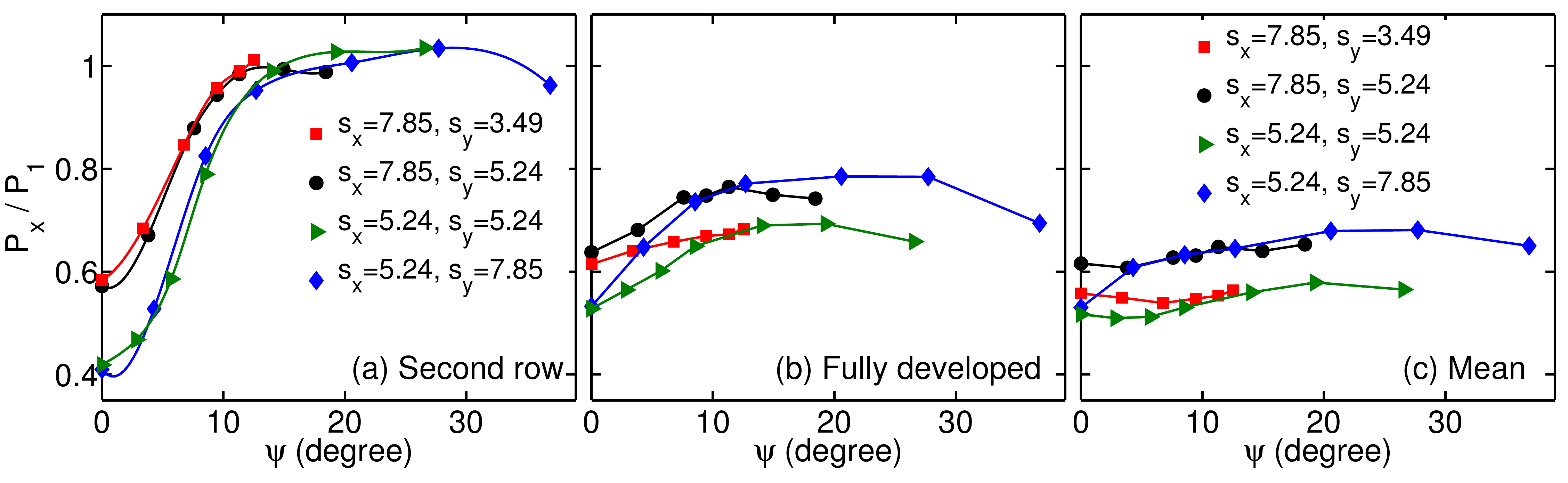}}
 \caption{Normalized power output as function of the alignment angle $\psi$ for (a) the second turbine row ($P_2 / P_1$) and (b) the fully developed region of the wind-farm ($P_\infty / P_1$). (c) Normalized mean turbine power output of all turbines in the farm $P_m/P_1$ as function of the alignment angle $\psi$.}
\label{figure5}
\end{figure}

In this section we investigate the influence of the alignment of the turbines with respect to the incoming flow in more detail. A detailed study of alignment effects on power output was presented in \cite{ste14b}, but for completeness, some results directly relevant to the present study are introduced here. The four cases discussed in this section correspond to cases A3, B3, C3 and E3 in table \ref{table1}. Figure \ref{figure5} compares the normalized power output in the second row and in the fully developed region of the wind-farm for these cases. It shows that the normalized power output in the second row, $P_2/P_1$ strongly depends on the relative placement of the turbines, which is indicated by the alignment angle $\psi$. This figure also shows that the turbine power output in the second row does not depend on $s_y$. The figure shows that for $s_x=7.85D$ the turbines in the second row are outside the wake of upstream turbines when the alignment angle is larger than about $10$ to $11$ degrees. For the smaller streamwise spacing of $5.24D$ this happens when $\psi$ is $13$ to $14$ degrees. We note that this compares well to the results in figure 8(a) of Hansen et al. \cite{han12}. That figure reveals that in Horns Rev an angle of about $11^\circ$ is necessary to ensure that turbines in the second row produce the same power as turbines in the first row. We note that figures \ref{figure2} and \ref{figure5}a show that for low $s_y$ the the normalized power output in the second row is slightly higher than that of the first row. This increase can be associated to a Venturi effect \cite{amm02,cha11b}.

Figure \ref{figure5}b shows the normalized power output with respect to turbines in the first row in the fully developed regime, $P_\infty/P_1$, for cases A3, B3, C3 and E3 in table \ref{table1}. For the cases in which the spanwise spacing is $5.24D$ and $3.49D$ the power output in the fully developed regime only weakly depends on the chosen alignment. This observation is consistent with the results for the aligned and staggered cases discussed above. However, for the larger spanwise spacing of $7.85D$ the power output in the fully developed regime is found to depend significantly on the alignment angle $\psi$. Figure \ref{figure5}b shows that the aligned case is the only configuration that significantly underperforms with respect to the best alignments. The power output for all other alignments is close to the power output of the staggered alignment. Here we emphasize that the staggered alignment does not necessarily give the highest power output, a point made in more detail in Ref. \cite{ste14b}. 

In figure \ref{figure5}c we show the normalized mean turbine power output for all turbines in the farm, $P_m/P_1$, as function of the alignment angle $\psi$. The figure shows that for most cases the highest mean power output is obtained for an intermediate alignment angle. Similar results were found in Refs.\ \cite{por13,ste14b}. The only exception is the wind-farm with a streamwise spacing of $7.85D$ and a spanwise spacing of $3.49D$. Here the staggered case gives the highest power output. The reason is that the highest power output for the entire wind-farm is obtained when the turbines in each subsequent row are placed such that they are not influenced by wake effects from turbines that are directly upstream~\cite{ste14b}. With the small spanwise spacing of $3.49D$ the influence of the wake is only minimized for the staggered configuration. 

\subsection{Vertical kinetic energy flux and wake recovery} \label{section_flux}
In the entrance region of the wind-farm, the capture rate of the kinetic energy from the wind at hub-height is the main determinant of the power output that can be obtained for a given turbine configuration. Therefore, in order to maximize power output in this region, the turbines need to be placed such that they are minimally affected by wakes from turbines directly upstream. In the fully developed regime, the power generated by the turbines mainly originates from the high velocity wind that is brought down to the hub-height plane by the vertical kinetic energy flux defined according to $\Phi= \overline{u^{\prime} w^{\prime}}\overline{u} $ \cite{cal10,cal10b}, where the overbar indicates time averaging. The vertical kinetic energy flux $\Phi$ is determined by a combination of flow phenomena, including the turbulence in the ABL, the increased turbulence levels caused by the wind-turbine wakes, and the cumulative growth of the internal wind-farm boundary layer \cite{cal10,cal10b}. Prior studies have examined how particular flow structures affect such vertical fluxes using analysis techniques such as quadrant analysis \cite{ham12} and proper orthogonal decomposition \cite{ver14}. Effects on scalar transport were also studied using wind tunnel experiments in Ref. \cite{mar12b}. From the LES we can determine the development of the vertical kinetic energy flux $\Phi$ in the wind-farm for each of the configurations considered in this study.

\begin{figure}
\centering
\subfigure{\includegraphics[width=0.49\textwidth]{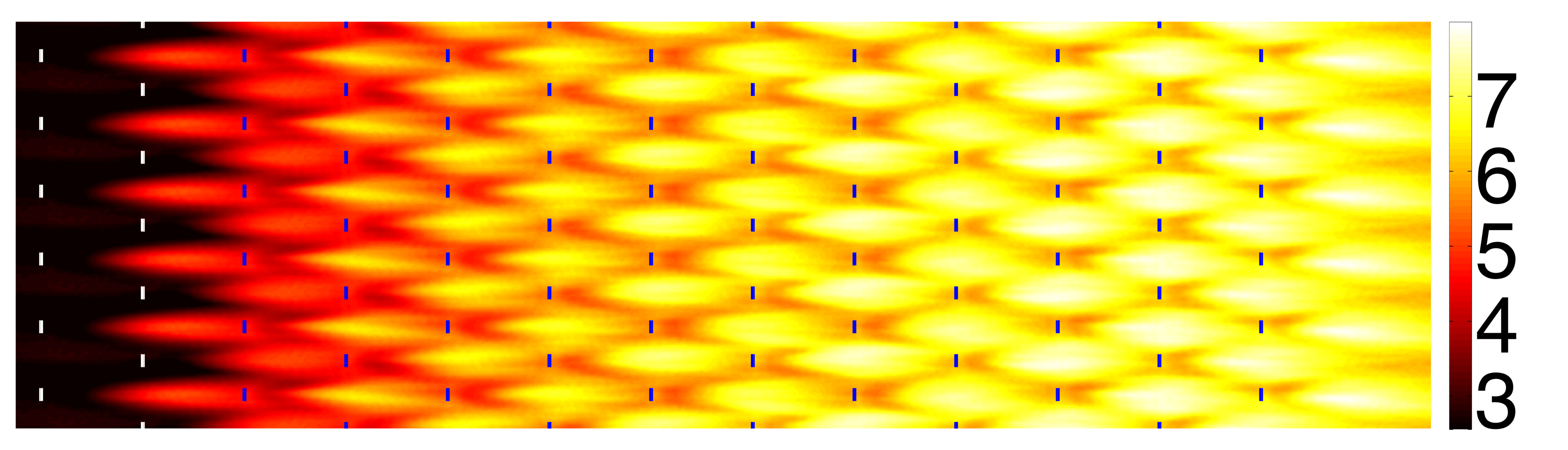}}
\subfigure{\includegraphics[width=0.49\textwidth]{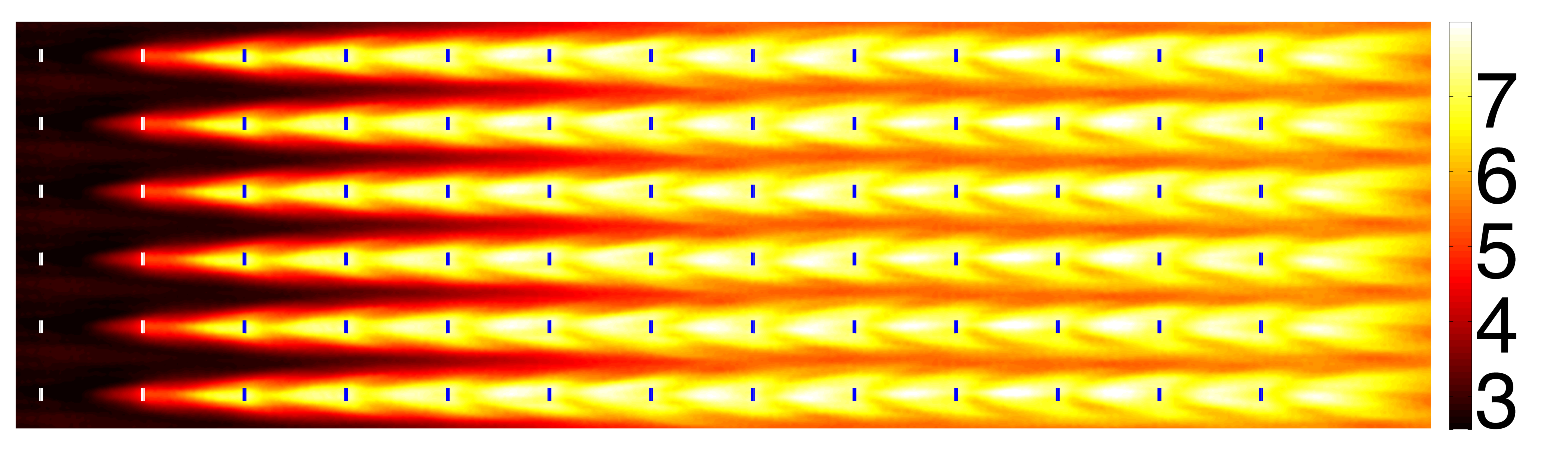}}
 \caption{Vertical kinetic energy flux $\Phi=\overline{ u^{\prime} w^{\prime}} \overline{u} $ at $z=2D$ normalized with $u_\mathrm{hub}^3/10^3$ for an  aligned (left) and a staggered (right) wind-farm with a streamwise spacing of $7.85D$ and a spanwise spacing of $5.24D$. The blue/white lines indicate the positions of the turbines.}
\label{figure6}
\end{figure}

\begin{figure}
\centering
\subfigure{\includegraphics[width=0.49\textwidth]{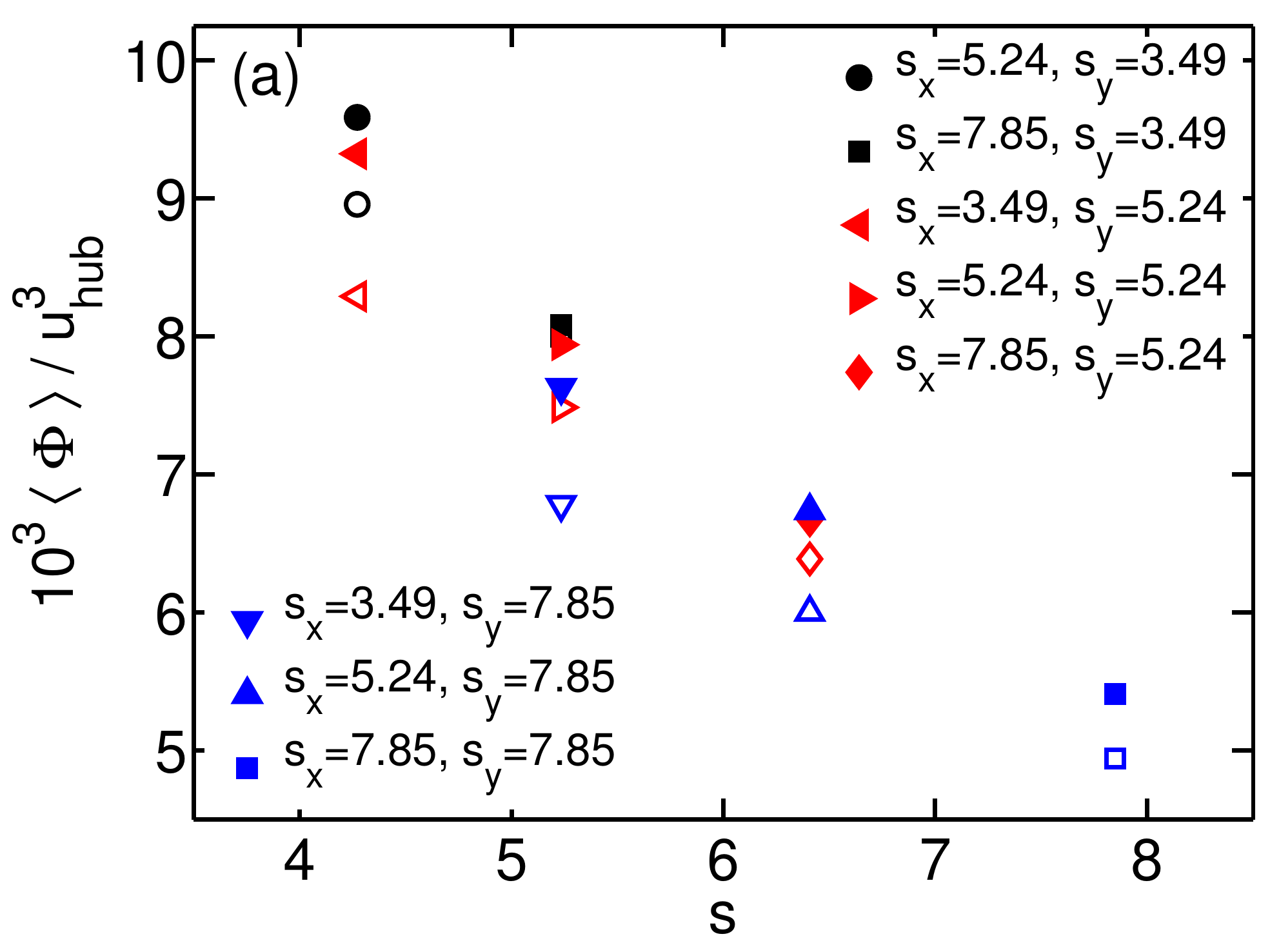}}
\subfigure{\includegraphics[width=0.49\textwidth]{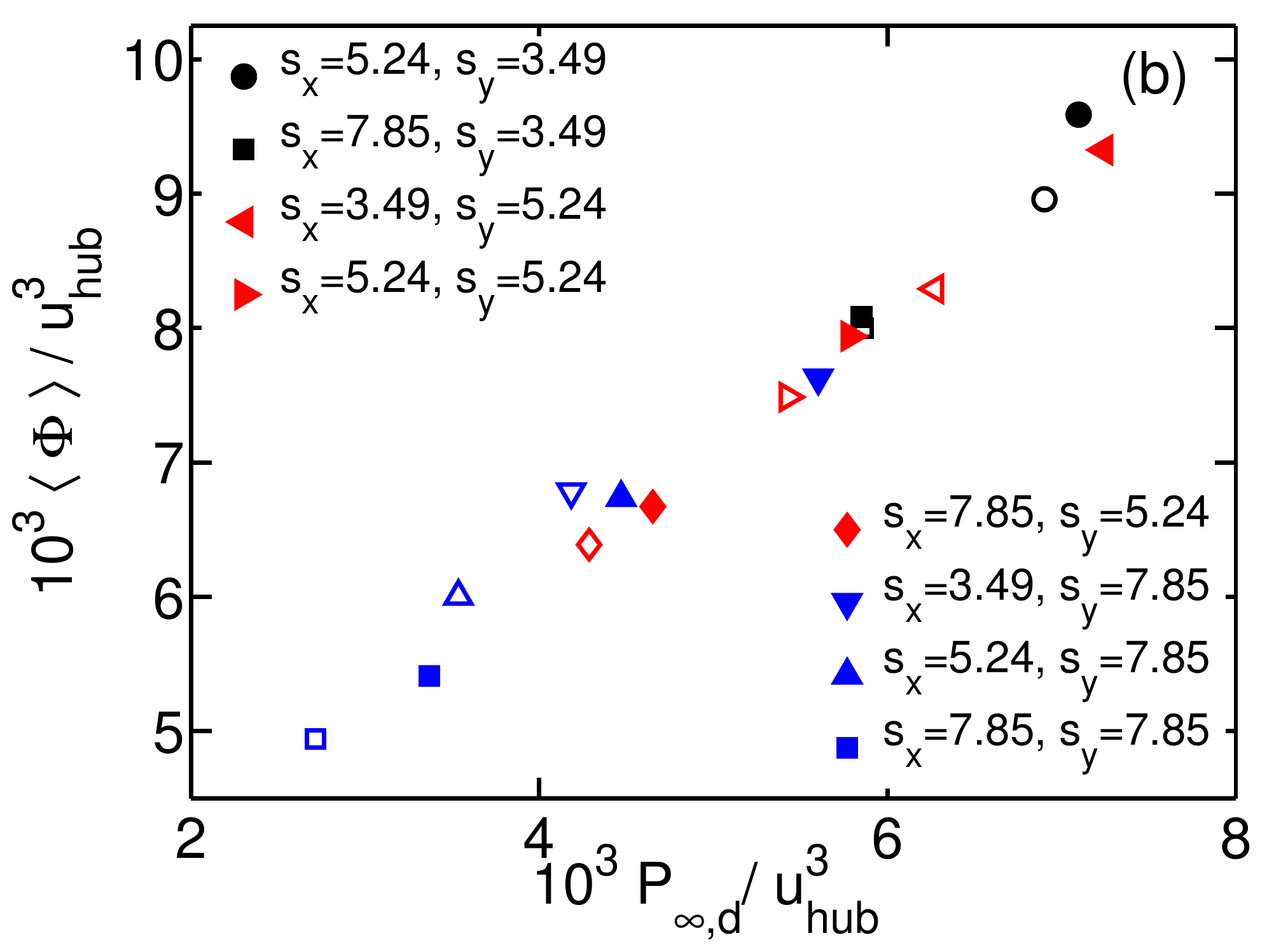}}
 \caption{Vertical kinetic energy flux $\langle \Phi \rangle = \langle \overline{ u^{\prime} w^{\prime}} \overline{u} \rangle$ as function of (a) the turbine spacing $s=\sqrt{s_xs_y}$ and (b) the power output density in the fully developed regime $P_{\infty,d}$. The results are normalized with the incoming hub-height velocity $u_\mathrm{hub}$.}
\label{figure7}
\end{figure}

Figure \ref{figure6} shows the vertical kinetic energy flux above the turbine plane at $z=2D$ for an aligned and a staggered wind-farm with a streamwise spacing of $7.85D$ and a spanwise spacing of $5.24D$. These plots show that the vertical kinetic energy flux becomes stronger with increasing downstream distance in the wind-farm. In addition, they indicate that the vertical kinetic energy flux is more localized for the aligned configuration than for the staggered one. Figure \ref{figure7} shows the average vertical kinetic energy as function of the turbine spacing, $s$, and as function of the power output density in the fully developed regime $P_{\infty,d}=1/2 c_\mathrm{ft}^\prime U_d^3$, where $U_d$ is the average disk velocity and $c_\mathrm{ft}^\prime =\pi C_T^\prime / (4s_xs_y)$ with $C_T^\prime=C_T/((1-a)^2)$ \cite{cal10,mey12}. The vertical kinetic energy flux and the power density are measured for streamwise distances between $6$ and $8$ km. The figure shows that there is a strong correlation between the vertical kinetic energy flux and the power density in the fully developed regime. This observation is in agreement with results reported in previous LES studies~\cite{cal10,ver14} and experiments \cite{cal10b,mar12b}.

\begin{figure}
\centering
\subfigure{\includegraphics[width=0.49\textwidth]{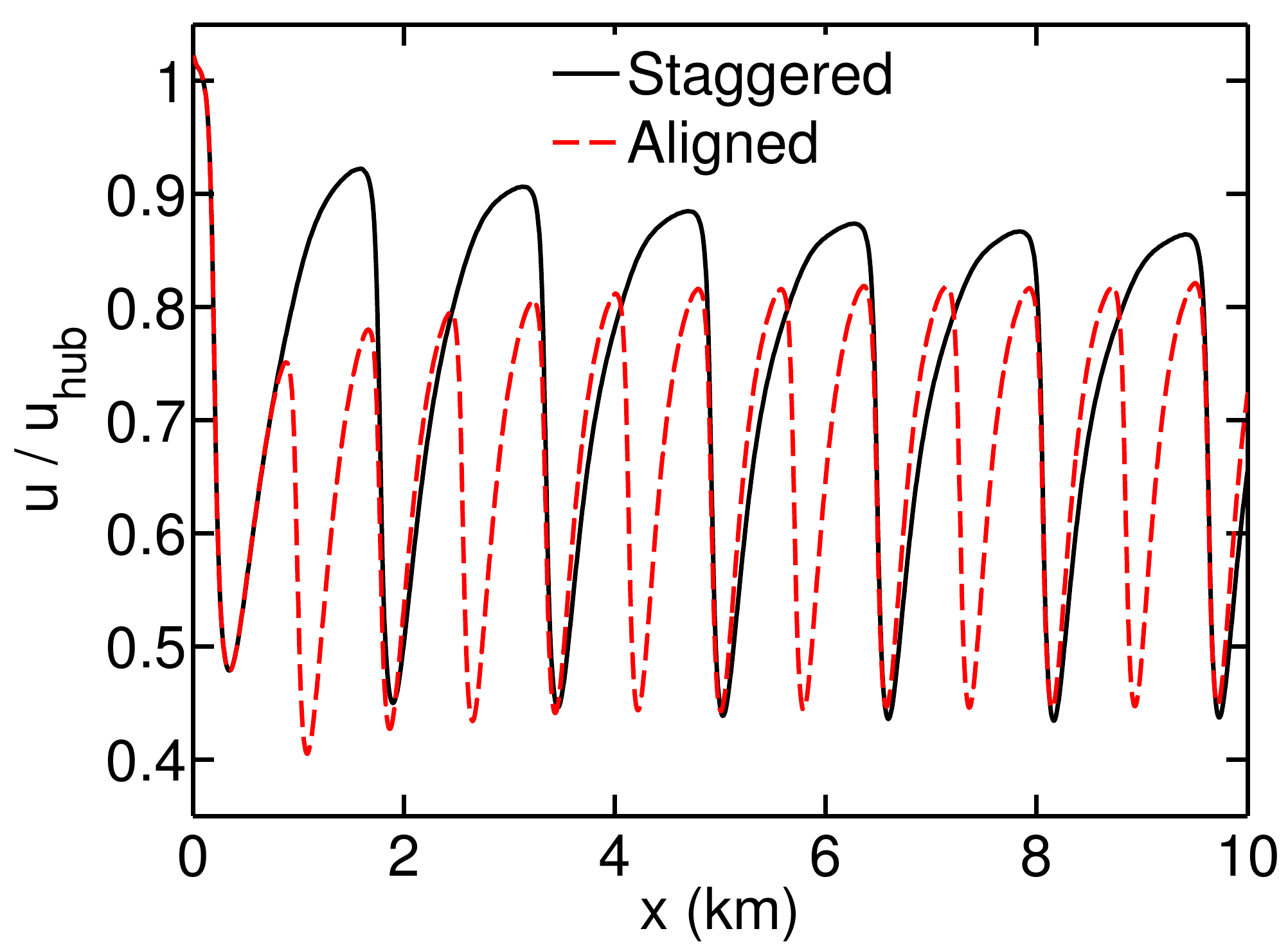}}
 \caption{Comparison of the average velocity in the center of the turbine columns (normalized by the incoming velocity at hub-heightht) for an aligned (dashed) and staggered (solid) wind-farm with a streamwise spacing of $7.85D$ and a spanwise spacing of $5.24D$.}
\label{figure8}
\end{figure}

The high velocity wind that is brought down to the hub-height plane due to the vertical kinetic energy flux is a direct and convenient indicator of how fast the wind-turbine wakes are recovering. The results in the previous section indicate that the vertical kinetic energy is more localized in aligned wind-farms than in staggered ones. Figure \ref{figure8} compares the average velocity at the turbine-nacelle line, i.e.\ at the height and spanwise location of the nacelle, as function of the downstream position for an aligned and a staggered configuration with a streamwise spacing of $7.85D$ and a spanwise spacing of $5.24D$. The figure shows that the wakes are recovering faster in the aligned configuration than in the staggered configuration. The wakes recover faster in the aligned wind-farms because of the strong localization of the vertical kinetic energy flux and the weaker lateral wake effects. However, as the available wake recovery length is twice as long for a staggered wind-farm than for an aligned wind-farm the power production is higher for the staggered case \cite{wu13}. Figure \ref{figure8} reveals that in the fully developed regime the velocity obtained about $7.5D$ behind a turbine (the peaks in the profiles for the aligned wind-farm) is significantly higher for an aligned configuration than the velocity at $7.5D$ behind the turbine when the turbines are staggered. This indicates that the underperformance of an aligned case compared to the staggered case is less than one might expect.

\section{Conclusions}
This study uses large eddy simulations (LES) to analyze power output and wake effects in large wind-farms. A number of different wind-farm configurations with various combinations of streamwise and spanwise turbine spacing are considered. The results show that in the fully developed regime the turbine power output for wind-farms arranged in a staggered configuration mainly depends on the geometric mean turbine spacing. However, for aligned wind-farms a sufficient streamwise spacing seems more important than the spanwise distance between the turbines. This observation has been made for streamwise and spanwise spacings in the range $[3.5,8]D$. For intermediate turbine alignments the power output in the fully developed regime seems to be roughly independent wind-farm of the alignment angle except for an (nearly) aligned configuration.

Analysis of the entrance region indicates that the wind-farm layout has a stronger influence in this region than in the fully developed regime. The power output in the entrance region of the wind-farm is mainly determined by the amount of energy that can be extracted from the incoming flow at hub-height. In the fully developed regime the vertical kinetic energy flux, which is a measure of the amount of high velocity wind that is brought down to the hub-height plane, is strongly correlated with power output density of the turbines. This quantity is more localized in an aligned configuration and therefore the wake recovery is faster for an aligned configuration than for a staggered configuration, however the power output of the staggered configuration is still generally larger. These results show that there are important differences between aligned and staggered wind-farms that need to be captured in wind-farm modeling tools. A recent model that is designed to capture these differences, is the the coupled wake boundary layer, CWBL, model introduced in \cite{ste14g,ste14h}.

{\it Acknowledgements:} This work is funded in part by the research program `Fellowships for Young Energy Scientists' (YES!) of the Foundation for Fundamental Research on Matter (FOM) supported by the Netherlands Organization for Scientific Research (NWO), and by the National Science Foundation through grant number 1243482 (the WINDINSPIRE project). This work used the Extreme Science and Engineering Discovery Environment (XSEDE), which is supported by the National Science Foundation grant number OCI-1053575 and the Cartesius and LISA cluster of SURFsara in the Netherlands.

\section*{Appendix: grid convergence analysis}
\begin{figure}[!t]
\begin{center}
\subfigure{\includegraphics[width=0.49\textwidth]{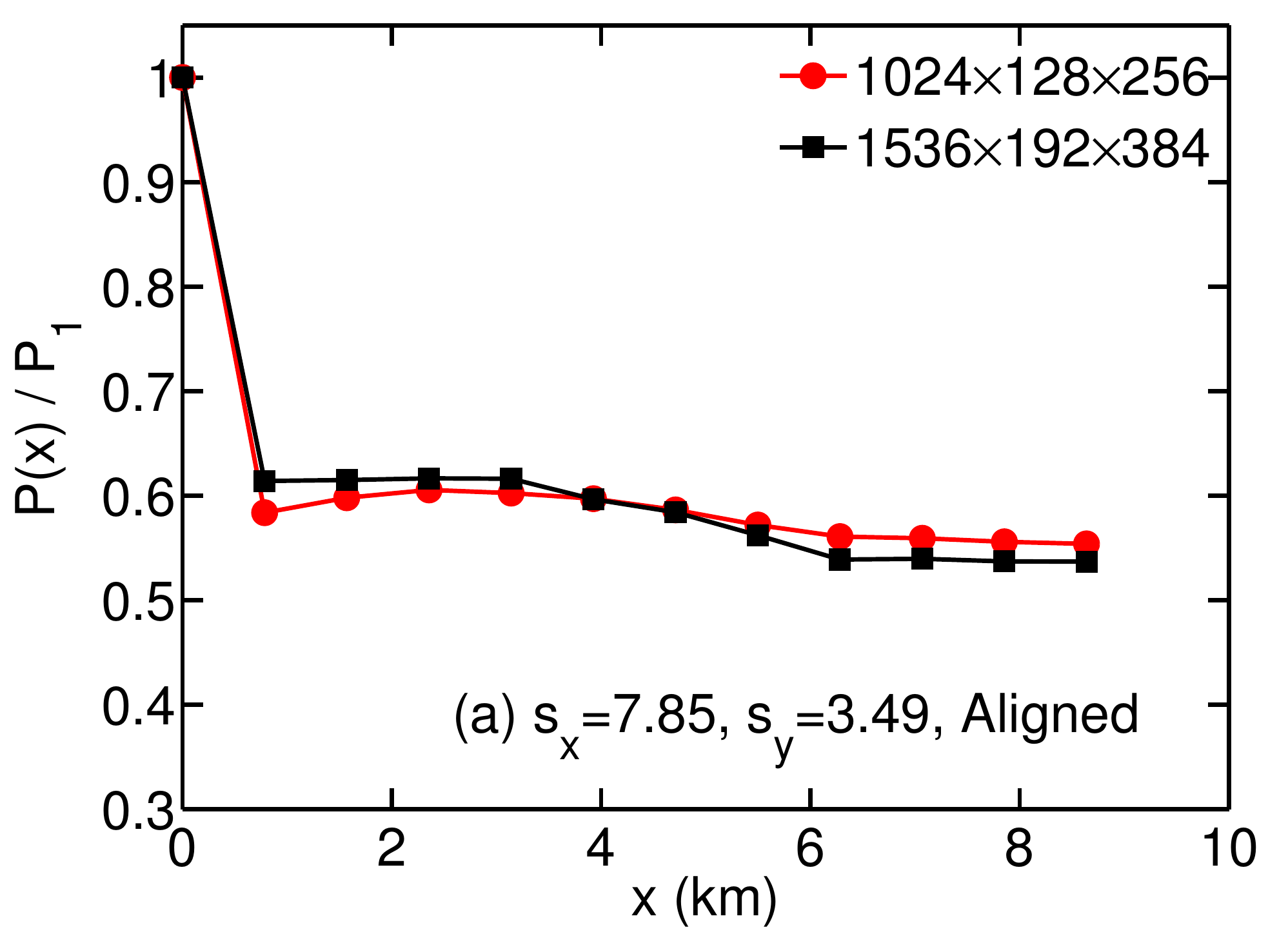}}
\subfigure{\includegraphics[width=0.49\textwidth]{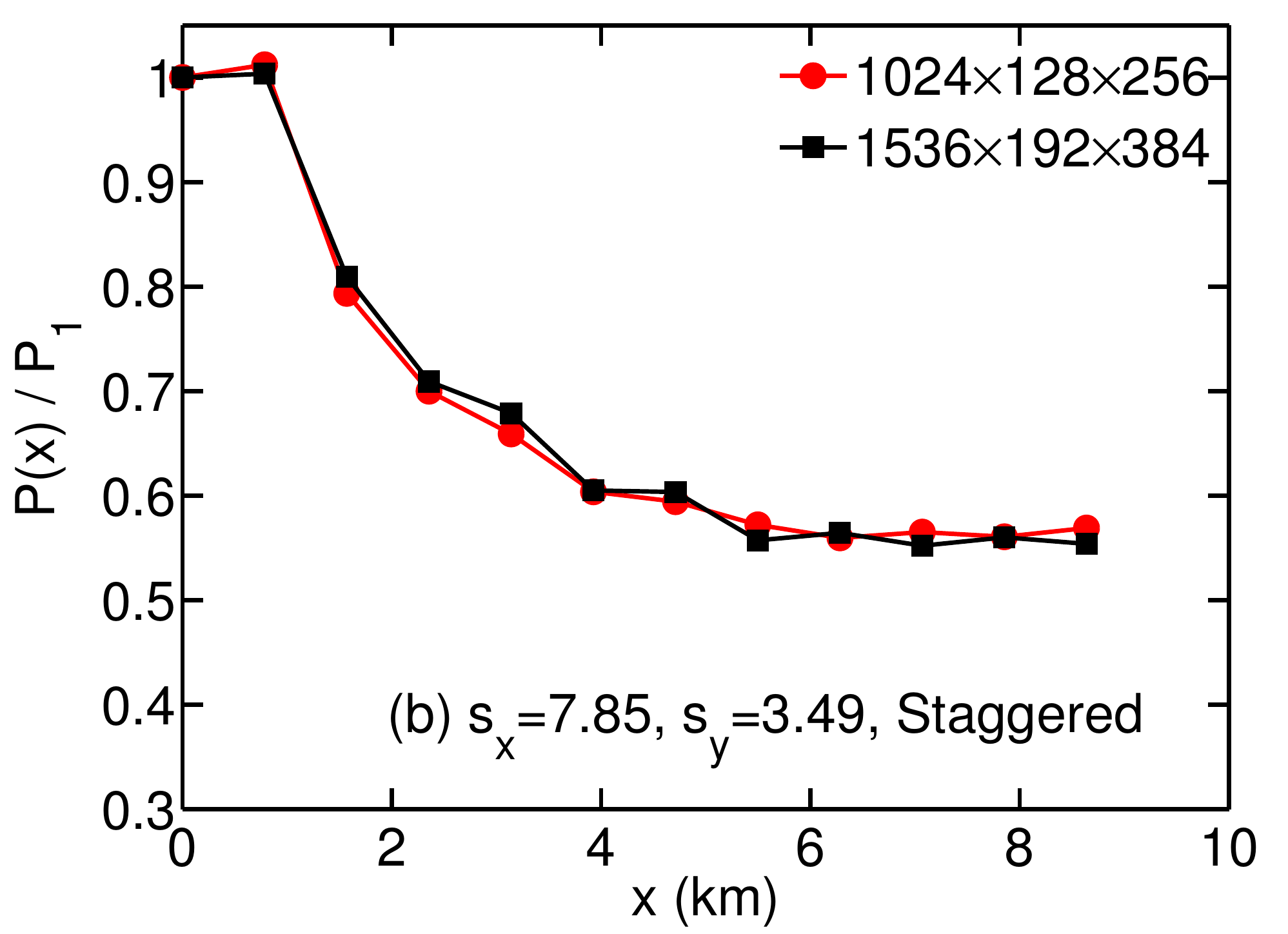}}
\subfigure{\includegraphics[width=0.49\textwidth]{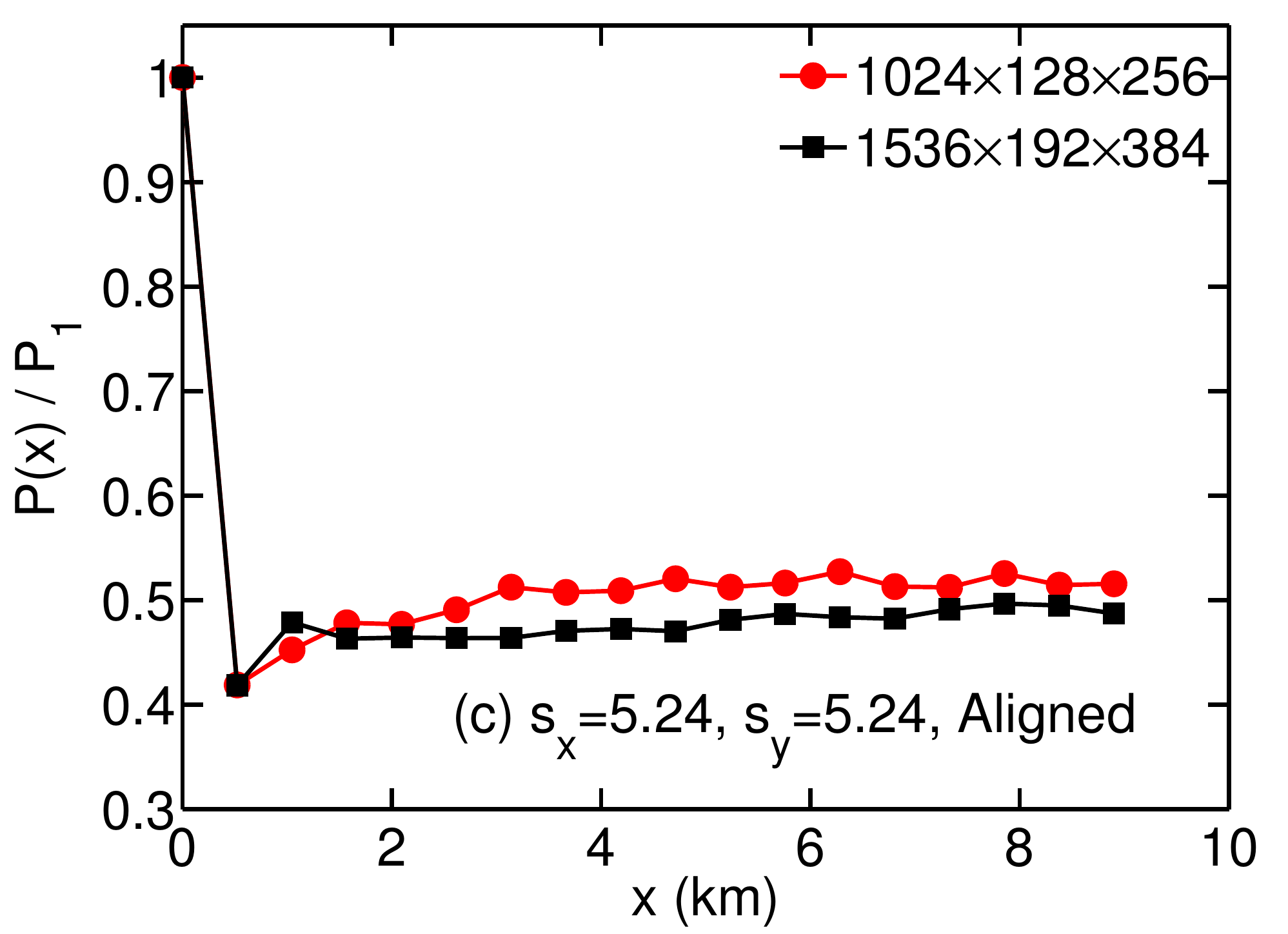}}
\subfigure{\includegraphics[width=0.49\textwidth]{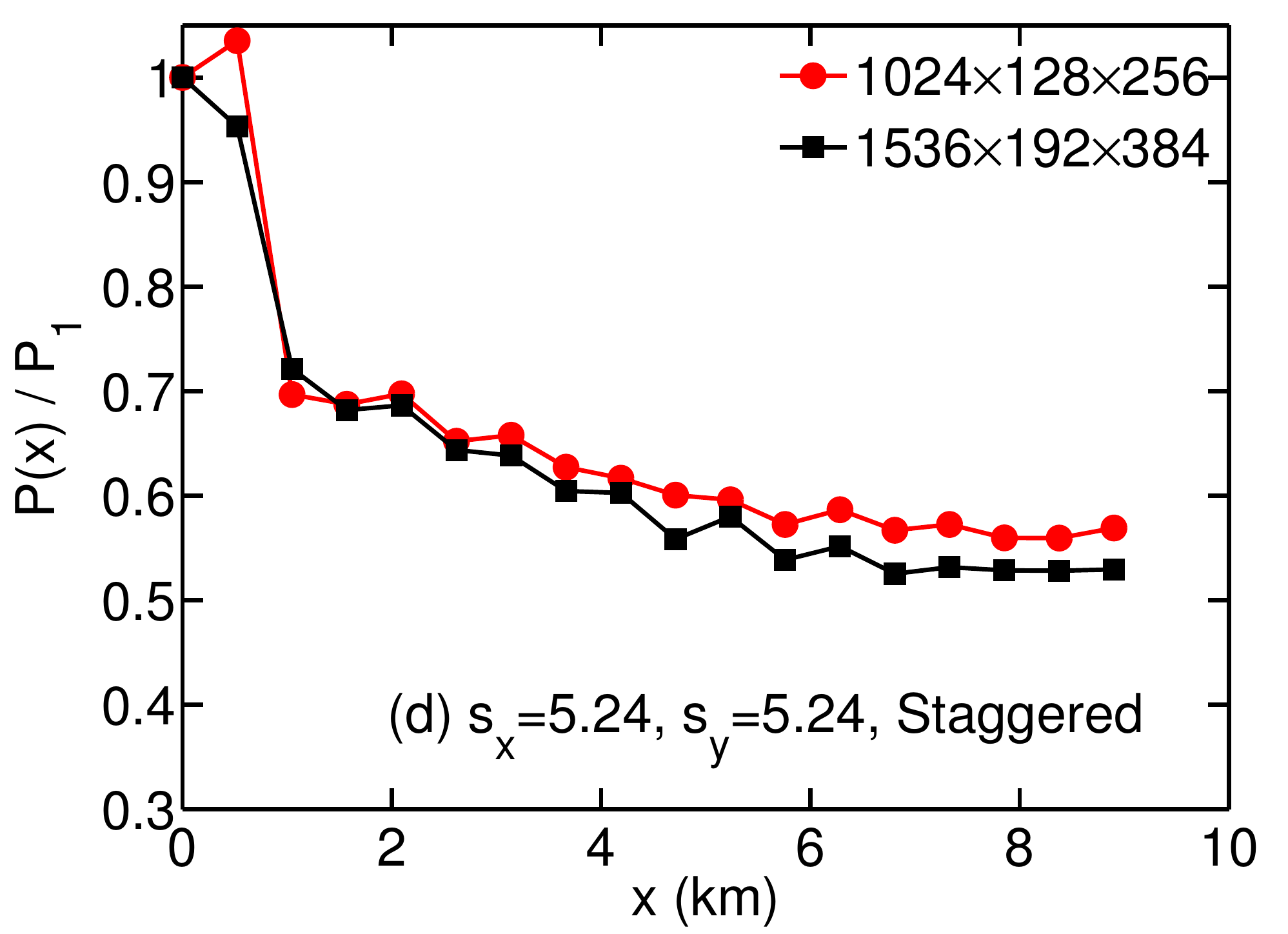}}
\caption{Comparison of the development of power output as function of downstream position for wind-farms with (a,b) a streamwise turbine spacing $s_x=7.85$ and a spanwise turbine spacing $s_y=3.49$ and (c,d) a streamwise turbine spacing $s_x=5.24$ and a spanwise turbine spacing $s_y=5.24$. The red and black data points give the results obtained using the $1024 \times 128 \times 256$ and $1536 \times 192 \times 384$ grids, respectively.}
\label{figure9}
\end{center}
\end{figure}

In this appendix we will compare the results of different resolution simulation, shown in table \ref{table1}, to verify that the the used numerical resolution gives sufficiently converged results. Figures \ref{figure9} shows a comparison of results obtained with LES simulations using $1024 \times 128 \times 256$ and $1536 \times 192 \times 384$ grids, respectively. In agreement with earlier results published in Ref.\ \cite{ste14} we find that there is a good agreement between the results obtained with these two grids when the streamwise spacing is $s_x=7.85$. The grid convergence is slightly less complete for the cases where the streamwise spacing is $5.24D$. It appears that with the shorter inter-turbine spacings the results are more sensitive to the resolution. We also note that the simulations with the different grid resolutions have been performed with different initial conditions. Considering the tendency of high and low velocity streaks to be in certain parts of the domain a relative shift of these structures with respect to the turbine locations can also be responsible for part of the observed differences. Based on these results, we can consider the grid convergence of sufficient quality for the purposes of the present study and for the trends identified in the main text.

\FloatBarrier


\end{document}